\documentclass[aps,pre,preprint,groupedaddress,showpacs,floatfix]{revtex4-1}
\usepackage{graphicx}
\usepackage{amsmath,amssymb,amsfonts,wasysym,color}
\usepackage{bm}

\begin{document}
\title{Fluctuation induced forces in  critical  films with disorder  at their surfaces}

\author{A.~Macio\l ek$^{1,2,3}$, O.~Vasilyev$^{1,2}$, V.~Dotsenko$^{4}$, and S.~Dietrich$^{1,2}$}
\affiliation{$^{1}$ Max-Planck-Institut f{\"u}r Intelligente Systeme,
  Heisenbergstra\ss e~3, D-70569 Stuttgart, Germany}
\affiliation{$^{2}$IV. Institut f\"ur Theoretische  Physik,
Universit\"at Stuttgart, Pfaffenwaldring 57, D-70569 Stuttgart, Germany }
\affiliation{$^{3}$Institute of Physical Chemistry, Polish Academy of Sciences,
Kasprzaka 44/52, PL-01-224 Warsaw, Poland}
\affiliation{$^{4}$ LPTMC, Universit\'e Paris VI - 75252 Paris, France}

\date{\today}

\begin{abstract}
We investigate the effect of quenched surface  disorder on effective interactions between two planar surfaces immersed in fluids 
 which are near
 criticality and belong to the Ising bulk universality class.  We consider the case that, in the absence of random surface fields,
the  surfaces of the film  belong  to the surface universality class of the so-called ordinary transition.
We find  analytically that in the linear weak-coupling regime, i.e., upon including the mean-field 
contribution  and Gaussian fluctuations, 
the presence of random surface  fields with zero mean  leads to  an  attractive, disorder-induced 
 contribution to the critical Casimir interactions
between the two confining surfaces. Our analytical, field-theoretic  results are compared with corresponding
Monte Carlo simulation data. 
\end{abstract}

\pacs{75.10.Nr, 64.60.an, 64.60.De, 68.35.Rh}

\maketitle

\section{Introduction}
\label{sec:intr}

Critical fluids  generate long-ranged forces
between their confining walls \cite{FdG}. This phenomenon   is an analogue of the  well-known Casimir effect in 
quantum electrodynamics \cite{Casimir,krech:99:0}. These so-called critical Casimir forces  (CCF)
are described in terms of universal scaling functions which are 
determined by the universality class of 
the bulk liquid and  the surface universality classes of the  confining surfaces \cite{binder}.
Classical fluids belong to the Ising bulk universality class. The  confining  
 surfaces, such as the container walls, typically  realize  the surface universality class of the
so-called normal  transition
 \cite{pershan,rafai,nature,nature_long,nellen}, which is characterized by
 a strong effective surface field acting on the order parameter  of the fluid.
For example, for a binary liquid mixture  near its demixing transition  the order parameter is defined as 
the deviation of the concentration from its critical value and 
the surface field describes the preference of the container wall
for one of the two components of the mixture. If there is no such preference,  the surface typically 
belongs to the surface universality class
of the  so-called ordinary transition 
corresponding to  Dirichlet boundary conditions (BC) for the order parameter \cite{binder}.
While Dirichlet BC are difficult to realize for classical fluids, 
they occur naturally for $^4$He near its superfluid transition  \cite{chan}.
For $^3$He/$^4$He mixtures near their tricritical point both types of BC can occur \cite{chan1}.
The scaling functions of the CCF for various bulk and surface universality classes have been determined analytically
 by mean field theory and beyond \cite{dietrich_krech,krech,upton,Borjan,md}  as well as 
by using  Monte Carlo
simulations \cite{VGMD,hucht,Hasenbusch};  if applicable  they are in  fine  agreement  with the experimental findings.

The  properties of the CCF $f_{\mathrm C}$, such  as the sign and the strength,  
depend crucially on the surface fields characterizing the confining surfaces.
By suitable surface treatments one can design  the sign of the surface fields, e.g.,  in the case of  aqueous mixtures  by fabricating 
hydrophilic or hydrophobic surfaces 
 \cite{nature,nature_long}.
One can also create   spatially varying  surface fields by modulating the chemical composition of the  surfaces.
In Ref.~\cite{NHB} a  smooth lateral variation of the surface field between  hydrophilic (positive surface field) and hydrophobic
(negative surface field) parts of the surface
has been achieved. Along this gradient, the CCF acting on  a colloidal particle   have  been 
measured. Various other crossover behaviors of CCF have  been analyzed
analytically and by computer simulations \cite{MMD,abraham_maciolek,vas-11,diehl}. The CCF for surfaces endowed
 with geometrically well defined alternating chemical stripes  have been investigated 
 experimentally and theoretically \cite{TZGVHBD}.
 
 Even  very carefully fabricated surfaces are not perfectly smooth or homogenous.
  Typically they  carry random chemical heterogeneities 
due to adsorbed impurities which act as local surface fields. Here we focus on  kinetically frozen  surface fields 
which   form quenched disorder and study  CCF acting in their presence. 
It is known  that for  quenched  random-charge disorder on surfaces of dielectric parallel walls at a distance $L$ 
 long-ranged forces  $ \propto \; L^{-2}$ emerge,  
even if the surfaces  are on average  neutral \cite{NDSHP,BAD}.  For large $L$ these 
forces, induced by quenched disorder, dominate  the pure van der Waals interactions, which decay as 
$L^{-3}$.
This differs from the behavior of systems which exhibit  quenched random surface fields (RSF). 

Recent {\it MC simulations} for   three-dimensional Ising films \cite{MVDD} have  shown that the presence
of random surfaces fields  with zero mean leads  to  CCF which  at bulk criticality asymptotically  decay
 as function of the film thickness $L$ 
as $L^{-3}$.  This  is the same behavior as for  the pure critical system and as for the pure van der Waals term.
This result has been obtained for the case  in which
 in the absence of RSF the  surfaces of the film  belong  to the surface universality class of
 the ordinary transition ($(o,o)$ BC). Roughly  speaking, such surfaces are realized in  systems in which droplets
 of, for example,  the demixed binary liquid mixture form a contact angle of $90^0$ with the chemically disordered substrate 
(see the intermediate substrate compositions  discussed in Ref.~\cite{NHB}).

It follows from  finite-size  {\it scaling analyses}, in agreement with the corresponding  {\it MC simulation} data \cite{MVDD}, 
 that for weak disorder  the CCF   still  exhibit scaling,  acquiring  a random field  scaling variable $w$ which  is zero for pure systems.
The data of  the  MC simulations suggest  that for weak disorder the difference 
between  the force corresponding to the  random surface field  
and the corresponding force for  the pure system (with  $(o,o)$ BC) varies as 
$f_{\mathrm C}(w\to 0)-f_{\mathrm C}(w=0)\sim w^2$. Moreover, for thin films such that $w \simeq 1$, 
 the presence  of RSF with vanishing mean value  increases 
significantly the strength   of CCF,  as  compared to  systems without them, and shifts the extremum  of the scaling function of $f_{\mathrm C}$   
 towards lower temperatures.   But $f_{\mathrm C}$  remains attractive. 
Finite-size scaling predicts that  asymptotically, for large $L$, $w$ scales as 
$w \sim L^{-0.26} \to 0$ indicating that this type of disorder
is an irrelevant perturbation of the  ordinary surface universality class.

This  conjecture  is consistent with results  of  Ref.~\cite{francesco}  in which  the so-called 'improved'
 Blume-Capel model~\cite{Blume,capel,Hasenbusch} was studied by MC simulations. This work  is concerned
with  quenched random disorder   which is present only at one of the two surfaces  
 and  is governed by the 
   binomial distribution, i.e., spins  at the surface, which are  subjected to disorder, 
 take  the value 1 with  probability $p$ and the value -1 with  probability $1-p$. It has been found that for $p=0.5$ 
 the  leading critical behavior of the CCF is still governed 
 by the ordinary fixed point.
These findings are in agreement with  the Harris criterion which concerns the relevance of disorder for bulk critical 
phenomena  and which has been generalized to  surface critical behavior
\cite{diehl_nusser1}. Within the framework and limitations of a weak-disorder expansion,
 quenched random  surface fields with vanishing mean value are   expected to be irrelevant if the pure
 system belongs to the ordinary surface universality class  \cite{diehl_nusser1}.
For the  three-dimensional ($d=3$) Ising model, in Ref.~\cite{mon_nig} 
this  was  pointed out and confirmed by Monte Carlo simulations.

For  semi-infinite systems the influence of random surface fields has been studied also in the context  of wetting (for reviews see Ref.~\cite{dietrich}) and surface
critical phenomena  \cite{mon_nig,diehl_nusser1,igloi,cardy}   (for a review see Ref.~\cite{Pleimling}).
In contrast to the case of simple fluids or binary  liquid mixtures,  for  complex fluids  surface disorder effects on  
Casimir-like interactions can be  dominant as shown recently for  nematic liquid-crystalline films  \cite{podgornik}.

So far, except of the general finite-size scaling analysis, the CCF  in the presence of RSF has not been studied  {\it analytically}.
This lack of  theoretical insight has rendered the corresponding MC simulations data obtained in Ref.~\cite{MVDD} rather difficult to interpret.
 Here we develop a fieldtheoretical approach   in terms of Gaussian perturbation theory, which is valid 
in the limit of weak disorder.   
As in   Ref.~\cite{MVDD}, we consider  films  of thickness $L$,  which 
in the three-dimensional bulk belong to the Ising universality class 
and  the surfaces  of which in the absence of RSF belong  to the surface universality class of the ordinary transition.

Our presentation is organized as follows. In Sec.~\ref{sec:sc} we briefly summarize  the results
of the finite-size scaling analysis in the presence of a random surface field, which were derived
in   Ref.~\cite{MVDD} and which form the analytical basis of the present study. 
In Sec.~\ref{sec:PS} we introduce  and discuss our  model in the absence of RSF.
In Sec.~IV we include  RSF  and calculate the corresponding scaling function of the CCF.
In Sec.~V we compare our findings with  MC simulations data and  provide an outlook. Technical details
of the calculations  in Sec.~IV are given in Appendices A and B.
\newpage

\section{Scaling}
\label{sec:sc}

Within  mean field  theory,  for pure systems   within the basin of attraction of the ordinary transition of semi-infinite 
systems,  in the ordered phase the order parameter profile exhibits  an extrapolation 
length $1/c$; $c=\infty$ is the fixed point of  the ordinary transition (o) ~\cite{binder}. 
Close to  this transition there is a single linear scaling field $g_{1}=H_1/{\tilde c}^{y_c}$
associated  with the dimensionless,  uniform surface 
field of strength $H_1$ and  with the  dimensionless surface enhancement parameter $\tilde c=ca$, 
where $a$ is a characteristic 
microscopic length scale of the system
  \cite{binder} such as the  amplitudes $\xi_0^{\pm}$ of the bulk correlation length
  $\xi_b(t=\frac{T-T_c}{T_c}\to 0^{\pm}) \;\simeq \; \xi_0^{\pm}|t|^{-\nu}$ 
  ( the symbol ``$\simeq$'' stands for asymptotic equality).
In the following  all lengths, such as $L$ and $1/c$, are taken in units of $a$ and thus are dimensionless.
The above scaling exponent is $y_c=\left(\Delta_1^{sp}-\Delta_1^{ord}\right)/\Phi$, 
 where  $\Delta^{ord}_{1}$ and $\Delta^{sp}_{1}$ are the surface counterparts at the ordinary and special transition, respectively, 
of the bulk gap exponent $\Delta$,
 and $\Phi$ is a  crossover exponent \cite{binder}.  Within mean field theory one has $y_c=1$ 
 whereas   $y_c(d=3)\;\approx \; 0.87 $ ~\cite{binder,GZ}.
 Close to the critical point, the {\it sing}ular part $f_{sing}$ of the free energy per $k_BT$ and per
 volume of a film of thickness $L$ scales as
 $f_{sing}(t,h_b,g_1;L^{-1}) \; \simeq  \; L^{-d}f_{sing}(L^{1/\nu}t,L^{\Delta/\nu}h_b,L^{\Delta^{ord}_1/\nu}g_1;1)$,
 where $h_b$ is  the dimensionless bulk ordering field.
 
In the presence of random surface fields with  a Gaussian distribution and  with the ensemble averages 
\begin{equation}
\label{eq:0}
\overline{H_{1}(\mathbf{r})}=0  \qquad \mathrm{and} \qquad \overline{H_{1}(\mathbf{r}) H_{1}(\mathbf{r}')} = h^{2}\delta(\mathbf{r} - \mathbf{r}'),
\end{equation}
where
$\mathbf{r}$ and $\mathbf{r'}$ denote dimensionless lateral positions,   finite-size scaling  predicts  ~\cite{MVDD}  that
the appropriate scaling variable, which  
replaces  $L^{\Delta^{ord}_1/\nu}g_1$ 
for the pure system,   is
\begin{equation}
 \label{eq:w}
 w \equiv  \kappa L^{\Delta^{ord}_1/\nu -(d-1)/2}h/c^{y_c} = \kappa L^{y_1-(d-1)/2}h/c^{y_c},
\end{equation}
 where $\kappa$ is a nonuniversal amplitude.
 The scaling exponent $y_1-(d-1)/2$ has been derived in Ref.~\cite{diehl_nusser1}; there 
 it was shown that it is related to  $\gamma_{11}$,  which is a standard surface susceptibility exponent of the ordinary transition:
 $\gamma_{11}=\nu(1-\eta_{||}) =-(d-1-2y_1)\nu$.
 In the MC simulation study reported in Ref.~\cite{MVDD} for the  three-dimensional $(d=3)$ Ising model, the following
 values of the critical exponents have been used:  $\Delta^{ord}_{1} \; \approx \;  0.46(2)$  \cite{GZ},  
$\Delta^{sp}_{1} \;  \approx \;  1.05$~\cite{binder}, 
$\Phi \;  \approx  \; 0.68$~\cite{binder}, 
 and $\nu \approx 0.63$ \cite{PV,Hasenbusch}.
These values  yield
$y_{1}-(d-1)/2 \;  \approx \; -0.26(6)$.
(More accurate estimates for the surface critical exponents at the  special and  ordinary 
transitions were  obtained recently 
from MC simulations \cite{Hasenbusch84}. They yield  $y_c\approx 1.282(5)$
and $y_1\approx 0.7249(6)$ so that $y_{1}-(d-1)/2  \approx -0.2750(4)$.)
Within    mean field theory, i.e., for $d=4$, one has $\Delta_1^{ord} = \nu=1/2$ \cite{binder} so that $y_{1}-(d-1)/2 = -1/2$.
Accordingly, for the  $d=3$ Ising model one has $w=\kappa (h/c^{0.87})L^{-0.26}$
whereas within mean field theory $w=\kappa (h/c)L^{-1/2}$.
 Because the scaling exponent  of  the random surface field is negative, the scaling field  $h/c^{y_c}$ is irrelevant in the  sense of
renormalization-group theory, which implies that for sufficiently  thick films the effect of  disorder
 is expected to be negligible.

\section{Pure system}
\label{sec:PS}

Within the  field-theoretic framework, near criticality a symmetric Ising film of thickness $L$ without ordering fields
is described by the (dimensionless) $d$-dimensional Ginzburg-Landau Hamiltonian
for the order parameter $\phi(\mathbf{r}, z)$ \cite{binder}:
\begin{equation}
 \label{1}
{\cal H}_{0}[\phi] = \int d^{d-1}r \int_{0}^{L}dz \Bigl[\frac{1}{2}\bigl(\nabla \phi\bigr)^{2} + \frac{1}{2}\tau\phi^{2}
+ \frac{1}{4!} g \phi^{4} +\frac{1}{2} c  \phi^{2} \bigl[\delta(z) + \delta(z-L)\bigr] \Bigr]
\end{equation}
where $\mathbf{r}$ is a $(d-1)$-dimensional lateral vector with $|\mathbf{r}| < R$;  the thermodynamic limit
requires $R \to \infty$, while the width $L$ remains large but finite. 
 In Eq.~(\ref{1}) and below the integral over $z$ is understood to be taken as $\lim_{\epsilon \to 0} \int_{0-\epsilon}^{L+\epsilon} dz$.
Negative values of the temperature variable
$\tau \sim t $
correspond to the bulk ferromagnetic phase which we study  in the following (concerning the disordered phase 
see Appendix B). We also assume that the surface coupling parameter
is large, i.e.,  $c \gg 1$, which corresponds to the ordinary transition
in semi-infinite systems.  In particular this implies that for  $\tau \geq 0$ the order parameter is identically zero.

The mean field equilibrium configuration $\phi_{*}(\mathbf{r},z)$ minimizes ${\cal H}_{0}[\phi]$,
satisfying $\phi_{*}''(z) = -|\tau| \phi_{*}(z) + \frac{1}{6} g \phi_{*}^{3}(z)$ with the boundary conditions
$\phi_{*}'(z)\Big|_{z=0} = c \phi_{*}(0)$ and $\phi_{*}'(z)\Big|_{z=L} = - c \phi_{*}(L)$.
With the bulk correlation length
$\xi_{-} = 1/\sqrt{2|\tau|}=\xi_{0}^{-}|t|^{-1/2}$ for $\tau < 0$ and
$\xi_{+} = 1/\sqrt{\tau}=\xi_{0}^{+}|t|^{-1/2}$ for $\tau > 0$ the function
$\phi_{*}(z,  t <0,L) = \phi_{0} \times \bigl(L/\xi_{0}^{-}\bigr)^{-\beta/\nu} \psi_{-}\bigl(z/L,L/\xi_{-}\bigr)$
decomposes into the amplitude $\phi_{0}$ of the bulk order parameter
$\phi_{b} = \phi_{0} |t|^{\beta}$, the power law $(L/\xi_{0}^{-})^{-\beta/\nu}$ and a universal scaling function
$\psi_{-}\bigl(s=z/L,x_{-}=L/\xi_{-}\bigr)$ with $0 \leq s \leq 1$ and
$\psi_{-}\bigl(1-s,x_{-}\bigr) = \psi_{-}\bigl(s,x_{-}\bigr)$;  $\phi_{*} \equiv 0$ for $t \geq  0$.
Within the present mean field theory (MFT)
$\tau = t/\bigl(2(\xi_{0}^{-})^{2}\bigr)$ and $\phi_{0} = \sqrt{3/g}/ \xi_{0}^{-}$
with the universal ratio $\xi_{0}^{+}/\xi_{0}^{-} = \sqrt{2}$. 
 The above scaling form for $\phi_{*}(z,t,L)$ holds beyond MFT.

The  MFT scaling function satisfies the differential equation
\begin{equation}
 \label{eq:diffeq}
 \frac{\partial^{2}}{\partial s^{2}} \psi_{-}(s,x_{-}) =- x_{-}^{2} \psi_{-}(s,x_{-})+\psi_{-}^{3}(s,x_{-})
\end{equation}
with the boundary conditions
$
\frac{\partial}{\partial s} \psi_{-}(s,x_{-})\Big|_{s=0} = c L  \psi_{-}(s=0,x_{-})
$
and
$
\frac{\partial}{\partial s} \psi_{-}(s,x_{-})\Big|_{s=1}~=~-c L  \psi_{-}(s=1,x_{-}) \, .
$
In the following we refrain  from indicating the dependence of the scaling function
$\psi_{-}$ on $x_{-}$ unless it is necessary.

The limit $c\to\infty$ has been studied in detail
in Ref.~\cite{Gambassi_Dietrich}. In this case the
scaling function $\psi_{-}(s)$ can be expressed
in terms of the Jacobi elliptic function $sn(s)$ which satisfies
$sn(s=0) = sn(s=1) = 0$
while its derivatives  at $s=0$ and at $s=1$ are nonzero.
This solution is the equilibrium one only for $\tau < \tau_{c} \equiv -\pi^{2}/L^{2}$;
for $\tau \geq \tau_{c}$ one has $\phi_{*}(z) \equiv 0$ 
{ (Beyond MFT this holds only for $\tau \geq 0$.
Within MFT, in the interval $-\pi^{2}/L^{2} < \tau \leq 0$, or
equivalently
$-\sqrt{2} \pi < x_{-}  \leq 0$, the film is disordered although the bulk is ordered.)
For large $x_{-}$ the scaling function $\psi_{-}(s)$ approaches that of
 the semi-infinite system:
$\psi_{-}\bigl(s\to 0, x_{-}\to\infty; s x_{-}=y_{-} \bigr) = x_{-}^{\beta/\nu} \, P_{-}\bigl(y_{-}=z/\xi_{-}\bigr)$
with $P_{-}(y_{-} = \infty) = 1$ and
$P_{-}(y_{-} \to 0) \sim y_{-}^{(\beta_{1}-\beta)/\nu}$
where $\beta_{1}(d=4) = 1$ and $\beta_{1}(d=3) = 0.80(1)$ \cite{diehl_rev,LZ} is a surface critical exponent;
within mean field theory $P_{-}(y_{-}) = \tanh(y_{-})$.
For large but finite values of the surface
enhancement parameter $cL$  the scaling function $\psi_{-}(s)$
is close to its fixed point form
corresponding to $c=\infty$ but still with nonzero values
 $\psi_{-}(0)$  and $\psi_{-}(1)$, in accordance with the boundary conditions
$\psi_{-}(s=0) = \psi_{-}(s=1) \sim 1/c$.

We now consider fluctuations $\varphi(\mathbf{r}, z)$ around the mean field equilibrium profile
$\phi_{*}(z) = \bigl(\phi_{0}\xi_{0}^{-}/L\bigr) \psi_{-}(z/L, L/\xi_{-}) \theta(-\tau)$,
 where $\theta$ is the Heaviside function.
Inserting $\phi(\mathbf{r}, z) = \phi_{*}(z) + \varphi(\mathbf{r}, z)$
into ${\cal H}_{0}[\phi]$
and subtracting the bulk contribution
${\cal H}_{0}[\phi_{b}] = S_{d-1} L \bigl(-\frac{3\tau^{2}}{2g}\bigr) \theta(-\tau)$
one obtains within  Gaussian approximation
\begin{eqnarray}
 \label{10}
{\cal H}_{0}[\varphi] - {\cal H}_{0}[\phi_{b}] &=& E_{0} S_{d-1}
+ \frac{1}{2} \int d^{d-1}r \int_{0}^{L}dz \Bigl[\bigl(\nabla \varphi\bigr)^{2}
+ \xi_{-}^{-2}\varphi^{2} - \xi_{-}^{-2} m_{-}(z/L, x_{-}) \varphi^{2} \\ \nonumber
&+& c  \varphi^{2} \bigl[\delta(z) + \delta(z-L)\bigr] \Bigr]
\end{eqnarray}
where
$
m_{-}(z/L, x_{-}) = \frac{3}{2} \Bigl[1 - \frac{1}{x_{-}^{2}}\psi_{-}^{2}(z/L)\Bigr], \; 
-\frac{1}{2} \xi_{-}^{2} = \tau,
$
$S_{d-1}$ is the $(d-1)$-dimensional  crossectional area of the system such that $S_{d-1} L$ is the volume of the film, and
$E_{0}$ is the mean field excess free energy density (per area) of a film over the bulk value (obtained by
inserting the mean-field profile $\phi_{*}(z)$ into Eq.~(\ref{1}) and subtracting ${\cal H}_{0}[\phi_{b}]$):
\begin{equation}
 \label{11}
E_{0} = -\frac{g}{24}  \int_{0}^{L}dz \; \phi_{*}^{4}(z)
+ L \frac{3\tau^{2}}{2g} \theta(-\tau) =
        -\frac{3}{8 g} L^{-3} \int_{0}^{1} ds \; \psi_{-}^{4}(s,L/\xi)
        + L \frac{3\tau^{2}}{2g} \theta(-\tau).
\end{equation}
 Note that $E_{0}$ depends on $c$ via $m_{-}$ and $\psi_{-}$. In the limit $L\to \infty$, $\; E_{0}$ reduces
to twice the surface energy of the corresponding semi-infinite system.
In terms of  the Fourier representation
\begin{equation}
\label{12}
\varphi(\mathbf{r}, z) = \int\frac{d^{d-1}p}{(2\pi)^{d-1}} \,
\frac{1}{L} \sum_{l=-\infty}^{+\infty} \tilde{\varphi}({\bf p}, l) \,
\exp\Bigl[i \, {\bf p} \cdot \mathbf{r} + \frac{2\pi i}{L} l z\Bigr] \; ,
\end{equation}
 where $\tilde{\varphi}({\bf p}, l)$ is given by the inverse Fourier transform
\begin{equation}
\label{12a}
\tilde{\varphi}({\bf p}, l) = \int d^{d-1}r \,\int_{0}^{L}dz \, 
\varphi(\mathbf{r}, z)  \,
\exp\Bigl[-i \, {\bf p} \cdot \mathbf{r} - \frac{2\pi i}{L} l z\Bigr] \; ,
\end{equation}
 Eq.~(\ref{10}) yields
\begin{equation}
  \label{14}
{\cal H}_{0}[\varphi] - {\cal H}_{0}[\phi_{b}]  = E_{0} S_{d-1} +
 \frac{1}{2} \int\frac{d^{d-1}p}{(2\pi)^{d-1}} \, \frac{1}{L^{2}} \sum_{l,l'=-\infty}^{+\infty}
G^{-1}_{l,l'}(p) \,
\tilde{\varphi}({\bf p}, l) \tilde{\varphi}(-{\bf p}, l')
\end{equation}
 where
\begin{equation}
 \label{15}
G^{-1}_{l,l'}(p) \; \equiv \; L \Bigl[p^{2} + \xi_{-}^{-2} + \frac{4\pi^{2}}{L^{2}} l^{2} \Bigr] \delta_{l, -l'}
                     \, - \,  \tilde{m}_{-}(l+l', x_{-}) \, + \, 2 c ;
\end{equation}
$\delta_{l, -l'}$ is the Kronecker symbol and (due to $\psi_{-}(s,x_{-}) = \psi_{-}(1-s,x_{-})$)
\begin{equation}
\label{16}
\tilde{m}_{-}(l,x_{-}) =
\frac{3x_{-}}{\xi_{-}} \int_{0}^{1/2} ds \, \bigl[1 - \frac{1}{x^{2}_{-}}\psi_{-}^{2}(s, x_{-})\bigr] \,
\cos\bigl( 2\pi s\, l \bigr) .
\end{equation}
Accordingly, one has  $\tilde{m}_{-}(l,x_{-}) =\frac{3}{\xi_{-} }\, f_{-}(l,x_{-})$
with
\begin{equation}
\label{16a}
f_{-}(l,x_{-}) \; = \; \int_{0}^{x_{-}/2} ds' \,
\bigl[1 - \frac{1}{x_{-}^{2}}\psi_{-}^{2}(s'/x_{-}, x_{-})\bigr]
\cos\Bigl( \frac{2\pi}{x_{-}} l s'\Bigr).
\end{equation}
Due to $\psi_{-}(s\ll 1, x_{-}\gg 1) \simeq x_{-} \tanh(s x_{-})$, taken to be valid up to $s=1/2$, one finds
\begin{equation}
f_{-}(l, x_{-} \gg 1) \simeq \int_{0}^{x_{-}/2} ds \bigl[1 - \tanh^{2}(s)\bigr] \cos\Bigl( \frac{2\pi l}{x_{-}}  s\Bigr)
 = 
\int_{0}^{x_{-}/2} ds \; \cosh^{-2}(s) \cos\Bigl( \frac{2\pi l}{x_{-}}  s\Bigr) \, .
\label{17} 
\end{equation}

In other words, the off-diagonal terms $\tilde{m}_{-}(l,x_{-})$ of the matrix given by  Eqs.~(\ref{15}) and (\ref{16})
can be approximated as follows:
\begin{equation}
 \label{17a}
 \tilde{m}_{-}(l,x_{-} \gg 1) \simeq  \frac{3}{\xi_{-}}  \int_{0}^{x_{-}/2} ds \cosh^{-2}(s) \cos\Bigl( \frac{2\pi l}{x_{-}} s\Bigr)
\, .
\end{equation}

\section{Random surface fields}
\label{sec:RSF}

Within the present model the presence of random surface fields is described by
\begin{equation}
 \label{18}
{\cal H}[\phi] = {\cal H}_{0}[\phi] +
\int d^{d-1}r \; \Bigl[ H_{1}(\mathbf{r}) \phi(\mathbf{r}, 0) +  H_{2}(\mathbf{r}) \phi(\mathbf{r}, L) \Bigr]
\end{equation}
where ${\cal H}_{0}[\phi]$ is the Ginzburg-Landau Hamiltonian of the pure system (Eq.~(\ref{1})) and
$H_{i}(\mathbf{r})$ ($i=1,2$) are random surface fields (see the Introduction).
$H_{1}$ and $H_{2}$ are taken to be uncorrelated.

Considering the fluctuations $\varphi(\mathbf{r}, z)$, as introduced in the  context of Eq.~(\ref{10}), leads to
\begin{equation}
\label{21}
{\cal H}[\varphi] = {\cal H}_{0}[\varphi] + \int d^{d-1}r \; \Bigl[ H_{1}(\mathbf{r}) \phi_{*}(0) +  H_{2}(\mathbf{r}) \phi_{*}(L)
+ H_{1}(\mathbf{r}) \varphi(\mathbf{r}, 0) +  H_{2}(\mathbf{r}) \varphi(\mathbf{r}, L) \Bigr]
\end{equation}
where ${\cal H}_{0}[\varphi]$ is the  Gaussian Hamiltonian of the pure system (Eq.~(\ref{14})).
The partition function is 
\begin{eqnarray}
 \nonumber
&& Z = \int {\cal D}[\varphi] \exp\Bigl\{-{\cal H}[\varphi]\Bigr\} \\
\nonumber
&& =
Z_{bulk} 
\int {\cal D}[\varphi] \exp\Biggl\{
- E_{0}S_{d-1} 
-  \frac{1}{2} \int\frac{d^{d-1}p}{(2\pi)^{d-1}} \, \frac{1}{L^{2}} \sum_{l,l'=-\infty}^{+\infty}
G^{-1}_{l,l'}(p) \, \tilde{\varphi}({\bf p}, l) \tilde{\varphi}(-{\bf p}, l')
\\
\nonumber
\\
&& -\int d^{d-1}r \; \Bigl[ H_{1}(\mathbf{r}) \phi_{*}(0) +  H_{2}(\mathbf{r}) \phi_{*}(L) \Bigr] 
- \int d^{d-1}r \; \Bigl[ H_{1}(\mathbf{r}) \varphi(\mathbf{r}, 0) +  H_{2}(\mathbf{r}) \varphi(\mathbf{r}, L) \Bigr]
\Biggr\}
 \label{21a}
\end{eqnarray}
where $Z_{bulk} = \exp\bigl\{- S_{d-1} L \bigl(-\frac{3\tau^{2}}{2g}\bigr) \theta(-\tau)\bigr\}$
and the elements $G_{l,l'}^{-1}(\bf{p})$ of the matrix $\hat{G}^{-1}(p)$
are given by  Eqs.~(\ref{15}) and (\ref{16}).
Regrouping the terms in the above equation one  finds
\begin{eqnarray}
 \nonumber
\frac{Z}{Z_{bulk}} &=& Z_{0} \exp\Biggl\{
- E_{0}S_{d-1} \, -  \, 
\int d^{d-1}r \; \Bigl[ H_{1}(\mathbf{r}) \phi_{*}(0) +  H_{2}(\mathbf{r}) \phi_{*}(L) \Bigr] 
\Biggr\}
\\
\nonumber
\\
&\times&
\Biggl< 
 \exp\Biggl\{
- \int d^{d-1}r \; \Bigl[ H_{1}(\mathbf{r}) \varphi(\mathbf{r}, 0) +  H_{2}(\mathbf{r}) \varphi(\mathbf{r}, L) \Bigr]
\Biggr\}
\Biggr>_{0} \; .
\label{21b}
\end{eqnarray}
Here $\langle ... \rangle_{0}$ denotes the {\it thermal} average taken with the
Gaussian Hamiltonian of the pure system (Eq.~(\ref{14})):
\begin{equation}
\label{21c}
\bigl< \bigl( ...\bigr) \bigr>_{0} \equiv 
Z_{0}^{-1}
\int {\cal D}[\varphi] \; (...) \; \exp\Biggl\{
-  \frac{1}{2} \int\frac{d^{d-1}p}{(2\pi)^{d-1}} \, \frac{1}{L^{2}} \sum_{l,l'=-\infty}^{+\infty}
G^{-1}_{l,l'}(p) \, \tilde{\varphi}({\bf p}, l) \tilde{\varphi}(-{\bf p}, l')
\Biggr\} \, .
\end{equation}
Using the general formula for Gaussian integrals,
\begin{equation}
\label{21e}
 \prod_{k=1}^{M} \int_{-\infty}^{+\infty} d\varphi_{k} \, 
\exp\Bigl\{ -\frac{1}{2} \sum_{k,k'=1}^{M} A_{k,k'} \, \varphi_{k} \varphi_{k'} \Bigr\} \; = \; 
(2\pi)^{M/2} 
\exp\Bigl\{ -\frac{1}{2} \mbox{Tr}\ln\bigl[\hat{A}\bigr] \Bigr\}
\end{equation}
which is valid for any matrix $A_{k,k'}$ with positive eigenvalues, one has
\begin{eqnarray}
 \nonumber
Z_{0} &=& \int {\cal D}[\varphi] \;  \exp\Biggl\{
-  \frac{1}{2} \int\frac{d^{d-1}p}{(2\pi)^{d-1}} \, \frac{1}{L^{2}} \sum_{l,l'=-\infty}^{+\infty}
G^{-1}_{l,l'}(p) \, \tilde{\varphi}({\bf p}, l) \tilde{\varphi}(-{\bf p}, l')
\Biggr\}
\\
\nonumber
\\
&=& 
{\cal B} \, \exp\Biggl\{
- \frac{1}{2} S_{d-1} \int
\frac{d^{d-1}p}{(2\pi)^{d-1}} \; \mbox{Tr}\ln\bigl[\hat{G}^{-1}(p)\bigr]
\Biggr\}
 \label{21d}
\end{eqnarray}
where $\mbox{Tr}$ denotes the matrix trace and the factor $L^{-2}$ in the exponential of Eq.~(\ref{21c}) is 
absorbed into the pre-exponential factor ${\cal B}$ in Eq.~(\ref{21d}). Note that the value of this
pre-exponential factor depends on the definition of the integration measure of the the fields $\varphi$. 
Since the prefactor ${\cal B}$ drops out of Eq.~(\ref{21c}) it is irrelevant for the considered problem 
 and thus  will be omitted in the further calculations.
The average in Eq.~(\ref{21b}) is calculated by using the Gaussian relation
$\langle\exp({\bf \lambda\cdot x})\rangle_{0} = 
\exp\bigl(\frac{1}{2}  \langle ({\bf \lambda\cdot x})^{2}\rangle_{0}\bigr)$.
Performing the Gaussian integrals over the fluctuating field $\varphi(\mathbf{r}, z)$
leads to
\begin{eqnarray}
 \nonumber
\ln\bigl( Z/Z_{bulk}\bigr) &=&
 - E_{0}S_{d-1} - \frac{1}{2} S_{d-1} \int
\frac{d^{d-1}p}{(2\pi)^{d-1}} \; \mbox{Tr}\ln\bigl[\hat{G}^{-1}(p)\bigr]
\\
\nonumber
\\
\nonumber
&-&
\int d^{d-1}r \; \Bigl[ H_{1}(\mathbf{r}) \phi_{*}(0) +  H_{2}(\mathbf{r}) \phi_{*}(L) \Bigr]
\\
\nonumber
\\
&+&
\frac{1}{2}
\Biggl< \Biggl(
\int d^{d-1}r
\Bigl[ H_{1}(\mathbf{r}) \varphi(\mathbf{r}, 0) +  H_{2}(\mathbf{r}) \varphi(\mathbf{r}, L) \Bigr] \Biggr)^{2}\Biggr>_{0}
\Biggr\} \, .
\label{22}
\end{eqnarray}
Note that the first two terms on the rhs of Eq.~(\ref{22}) are independent of $H_{1}$ and $H_{2}$.
Accordingly, for the free energy (per $k_{B} T_{c}$ and in excess of the bulk contribution
$\overline{F}_{b}$) {\it averaged over the random surface fields}
we find ($\overline{H_{1} \; H_{2}} \; = \;  \overline{H_{1}} \; \overline{H_{2}} \; = \; 0$)
\begin{equation}
 \label{23}
\frac{\overline{F}-\overline{F}_{b} }{S_{d-1}} =
E_{0}  + \frac{1}{2} \, \int
\frac{d^{d-1}p}{(2\pi)^{d-1}} \; \mbox{Tr}\ln\bigl[\hat{G}^{-1}(p)\bigr]
- \frac{1}{2} h^{2}
\Bigl(\bigl<\varphi^{2}(\mathbf{r}, 0)\bigr>_{0} + \bigl<\varphi^{2}(\mathbf{r}, L)\bigr>_{0}\Bigr) \, .
\end{equation}
 In terms of the Gaussian integral, Eqs.~(\ref{21c}) and (\ref{21d}), for the correlation function of the 
fields $\tilde{\varphi}({\bf p}, l)$ one obtains
\begin{equation}
 \bigl<\tilde{\varphi}({\bf p}, l) \tilde{\varphi}({\bf p}', l')\bigr>_{0} \; = \; 
L^{2} G_{l,l'}(p) \; (2\pi)^{d-1}\delta\bigl({\bf p} + {\bf p}'\bigr) \, .
\end{equation}
Thus, using the Fourier representation in Eq.(\ref{12}) the thermal averages 
in Eq.~(\ref{23}) can be represented as
\begin{eqnarray}
 \nonumber
&&\bigl<\varphi^{2}(\mathbf{r}, L)\bigr>_{0} = \bigl<\varphi^{2}(\mathbf{r},0)\bigr>_{0}
\\
\nonumber
\\
\nonumber
&&= \int \frac{d^{d-1}p}{(2\pi)^{d-1}} \int \frac{d^{d-1}p'}{(2\pi)^{d-1}}\,
\frac{1}{L^{2}}\sum_{l,l'=-\infty}^{+\infty} 
\bigl< \tilde{\varphi}({\bf p}, l) \,\tilde{\varphi}({\bf p'}, l') \bigr>_{0}
\exp\Bigl[i \, ({\bf p} + {\bf p}') \cdot \mathbf{r} \Bigr]
\\
\nonumber
\\
\nonumber
&& =
\int \frac{d^{d-1}p}{(2\pi)^{d-1}} \int \frac{d^{d-1}p'}{(2\pi)^{d-1}}\,
\frac{1}{L^{2}}\sum_{l,l'=-\infty}^{+\infty}
L^{2} \; G_{l,l'}(p) \;(2\pi)^{d-1}\delta\bigl({\bf p} + {\bf p}'\bigr)
\exp\Bigl[i \, ({\bf p} + {\bf p}') \cdot \mathbf{r} \Bigr]
\\
\nonumber
\\
&& =
\int \frac{d^{d-1}p}{(2\pi)^{d-1}} \; \sum_{l,l'=-\infty}^{+\infty} G_{l,l'}(p)
\label{24}
\end{eqnarray}
where $\hat{G}(\bf{p})$ is defined via $\hat{G}^{-1}(\bf{p})$ as
$\sum_{l''=-\infty}^{+\infty} G^{-1}_{l,l''}({\bf p}) G_{l'',l'}({\bf p}) = \delta_{l,-l'}$.
Within the present approach, 
$ \bigl<\varphi^{2}(\mathbf{r}, L)\bigr>_{0} = {\cal E}(z=L)$ 
and $ \bigl<\varphi^{2}(\mathbf{r},0)\bigr>_{0}={\cal E}(z=0)$, 
where ${\cal E}(z=L)={\cal E}(z=0)$ 
 is the fluctuation contribution to the energy density  at the surfaces of the pure film system
without surface fields \cite{DD,KED};
this quantity is independent of $\mathbf{r}$.

Subtracting the free energy of the pure system, one has
for the free energy contribution $\Delta F(h, L)$ due to the random field:
\begin{equation}
 \label{25}
\frac{\Delta F(h, L)}{S_{d-1}} =
-  h^{2} \bigl<\varphi^{2}(\mathbf{r}, 0)\bigr>_{0} =
- h^{2}
\int\frac{d^{d-1}p}{(2\pi)^{d-1}} \,\sum_{l,l'=-\infty}^{+\infty} G_{l,l'}(p)
.
\end{equation}
In order to deal with the divergent integral over $p$ we use dimensional regularization.
Using the explicit expressions in Eqs.~(\ref{15}) and ({\ref{17a}) together with the relation  \linebreak
$\sum_{l''=-\infty}^{+\infty} G^{-1}_{l,l''}(p) G_{l'',l'}(p)~=~ \delta_{l, -l'}$, one finds
(see Appendix A)
\begin{equation}
 \label{26}
 \sum_{l,l'=-\infty}^{+\infty} G_{l,l'}(p) \; = \;
\frac{g(p,L,\xi_{-})}{
1 + 2c \, g(p,L,\xi_{-}) - g_{1}\bigl((p\xi_{-})^{2} , x_{-}\bigr) }
\end{equation}
where
\begin{eqnarray}
 \nonumber
g(p,L,\xi_{-}) &=& \frac{1}{L}
\sum_{l=-\infty}^{+\infty} \Bigl[p^{2} + \xi_{-}^{-2} + \Bigl(\frac{2\pi l}{L}\Bigr)^{2}\Bigr]^{-1} \; = \;
\frac{1}{2\sqrt{p^{2}+\xi_{-}^{-2}} \, \tanh\Bigl(\frac{L}{2} \sqrt{p^{2}+\xi_{-}^{-2}}\Bigr)} \; = \;
\\
\nonumber
\\
&=&
\frac{\xi_{-}}{2\sqrt{1+ (p\xi_{-})^{2}} \, \tanh\Bigl(\frac{x_{-}}{2} \sqrt{1+ (p\xi_{-})^{2}}\Bigr)}
\label{27}
\end{eqnarray}
and 
\begin{equation}
g_{1}\bigl((p\xi_{-})^{2} , x_{-}\bigr) \; = \; 
\frac{3}{2\sqrt{(p\xi_{-})^{2} + 1} } \int_{0}^{x_{-}/2} ds \; \cosh^{-2}(s) \; \exp\Bigl\{- s \sqrt{(p\xi_{-})^{2} + 1} \Bigr\} \; .
\label{28}
\end{equation}
\begin{figure}[h]
\includegraphics[scale=1]{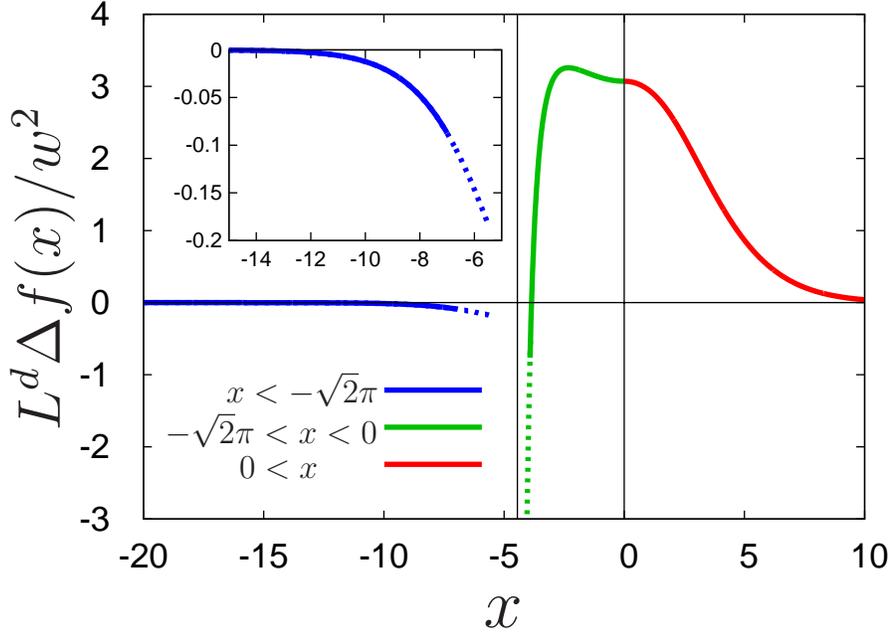}
\caption{The scaling function $\Delta f L^{d}/w^{2}$ of the contribution to the critical Casimir force
due to random surface fields as given by Eqs.~(\ref{33}) and (\ref{34}).
The underlying Gaussian approximation is evaluated for $d=4$, for which it is exact.
The scaling variable $x$ equals $-L/\xi_{-}$ for $T \leq T_{c, bulk}$ and $+L/\xi_{+}$ for $T\geq T_{c, bulk}$.
Withing mean field theory the critical temperature $T_{c, film}(L)$ corresponds to $x_{c} = -\sqrt{2} \, \pi$.
The left vertical  line corresponds to $x = x_{c} = -\sqrt{2}\pi$
whereas the right vertical line denotes $x=0$, i.e., the bulk critical point. The inset shows the magnified 
part of $\Delta f L^{d}/w^{2}$ as it is given by Eq.~(\ref{33}).} 
\label{fig:1}
\end{figure}

 By inserting  Eq.~(\ref{26}) into Eq.~(\ref{25}) and rearranging  the integrand one obtains
\begin{equation}
\label{29}
\frac{\Delta F}{S_{d-1}} \; = \; \frac{h^{2}}{2c} \int \frac{d^{d-1}p}{(2\pi)^{d-1}} \Biggl(\frac{1-g_{1}}{2c g + 1 -g_{1}} \; - \; 1\Biggr)
\; .
\end{equation}
 In the next step, we insert  the explicit expressions for $g$ and $g_{1}$  (Eqs.~(\ref{27}) and (\ref{28})) and 
determine  the surface  terms by taking 
  the limit $x_{-} =  L/\xi_{-} \to \infty$. Subtracting these  $L$-independent terms   
we obtain  the  excess free energy (denoted  by $\Delta\tilde{F}$)
\begin{eqnarray}
 \nonumber
\frac{\Delta \tilde{F}}{S_{d-1}} &=& \frac{h^{2}}{2c^{2}}
\int \frac{d^{d-1}p}{(2\pi)^{d-1}}
\sqrt{p^{2}+\xi_{-}^{-2}} \, \tanh\Bigl(\frac{L}{2} \sqrt{p^{2}+\xi_{-}^{-2}}\Bigr) \times
\\
\nonumber
\\
&\times&
\Biggl[
1 \; - \; 
\frac{3}{2\sqrt{(p\xi_{-})^{2} + 1} } \int_{0}^{x_{-}/2} ds \; \cosh^{-2}(s) \; \exp\Bigl\{- s \sqrt{(p\xi_{-})^{2} + 1} \Bigr\}
\Biggr] \, .
 \label{30}
\end{eqnarray}
This expression is valid for large $c$   to  leading order in   an expansion in terms of $1/c$. 
Using the substitution $p = y/\xi_{-}$ and integrating over the angular part of the momenta, we obtain
\begin{eqnarray}
\nonumber
\frac{\Delta \tilde{F}}{S_{d-1}} &=&
\frac{ \pi^{\frac{1-d}{2}} }{\Gamma\bigl(\frac{d+1}{2}\bigr) } \,
\frac{h^{2}}{2c^{2}\xi_{-}^{d}}  \;
\int_{0}^{\infty} dy \; 
y^{d-2} \sqrt{y^2 + 1} 
\tanh\Bigl(\frac{x_{-}}{2} \sqrt{1+ y^{2}}\Bigr) \times
\\
\nonumber
\\
&\times&
\Biggl[
1 \; - \; 
\frac{3}{2\sqrt{y^{2} + 1} } \int_{0}^{x_{-}/2} ds \; \cosh^{-2}(s) \; \exp\Bigl\{- s \sqrt{y^{2} + 1} \Bigr\}
\Biggr] \, .
\label{31}
\end{eqnarray}
Taking the negative derivative of this expression with respect to $L$,  which 
amounts to
$-\frac{\partial}{\partial L} = -\xi_{-}^{-1} \frac{\partial}{\partial x_{-}}$, 
renders the critical Casimir force $\Delta f$, per $k_{B}T$ and per area $S_{d-1}$,
in excess to its value without  random fields:
\begin{eqnarray}
 \nonumber
\Delta f &\simeq&  
-\frac{ \pi^{\frac{1-d}{2}} }{\Gamma\bigl(\frac{d+1}{2}\bigr) } \,
\frac{h^{2}}{4c^{2}\xi_{-}^{d+1}} \\ \nonumber
&& \int_{0}^{\infty} dy \; y^{d-2}
\Biggl\{
\frac{y^2+1}{\cosh^{2}\Bigl[\frac{x_{-}}{2} \sqrt{1+ y^{2}}\Bigr]} 
\Biggl[
1 - \frac{3}{2\sqrt{y^{2} + 1} } \int_{0}^{x_{-}/2} ds 
\frac{\exp\Bigl(- s \sqrt{y^{2} + 1} \Bigr)}{\cosh^{2}(s)}
\Biggr]
\; - \; 
\\
\nonumber
\\
&-& 
\frac{3\tanh\Bigl(\frac{x_{-}}{2} \sqrt{1+ y^{2}}\Bigr) \exp\Bigl(- \frac{1}{2} x_{-} \sqrt{y^{2} + 1} \Bigr)}{
2 \cosh^{2}\bigl(\frac{1}{2} x_{-}\bigr)}
\Biggr\} \, .
\label{32}
\end{eqnarray}
Replacing $\xi_{-}$ by $ L/x_{-} $ and identifying the dimensionless scaling variable
$w^{2} = h^{2}/(c^{2} L)$ (see Introduction),
  leads to the following final result:
\begin{eqnarray}
\Delta f &\simeq&  
-\mathcal{A}(d)
\frac{w^{2} x_-^{d+1}}{L^{d}}
\int_{0}^{\infty} dy \; y^{d-2} \times \nonumber
\\
\nonumber
&\times&
\Biggl\{
\frac{y^2+1}{\cosh^{2}\Bigl[\frac{x_{-}}{2} \sqrt{1+ y^{2}}\Bigr]} 
\Biggl(
1 - \frac{3}{2\sqrt{y^{2} + 1} } \int_{0}^{x_{-}/2} ds 
\frac{\exp\Bigl[- s \sqrt{y^{2} + 1} \Bigr]}{\cosh^{2}(s)}
\Biggr)
\\
\nonumber
\\
&-& 
\frac{3\tanh\Bigl(\frac{x_{-}}{2} \sqrt{1+ y^{2}}\Bigr) \exp\Bigl[- \frac{1}{2} x_{-} \sqrt{y^{2} + 1} \Bigr]}{
2 \cosh^{2}\bigl(\frac{1}{2} x_{-}\bigr)}
\Biggr\}
\label{33}
\end{eqnarray}
which is valid for $x_{-} \gg 1$ , $c\gg 1$ (to leading order $O(1/c^3)$; compare Eqs.~(\ref{16a}) and (\ref{17})). 
The prefactor   is given by $\mathcal{A}(d) = \pi^{\frac{1-d}{2}}/(4\Gamma\bigl(\frac{d+1}{2}\bigr))$ .

\begin{figure}[h]
   \includegraphics[scale=1]{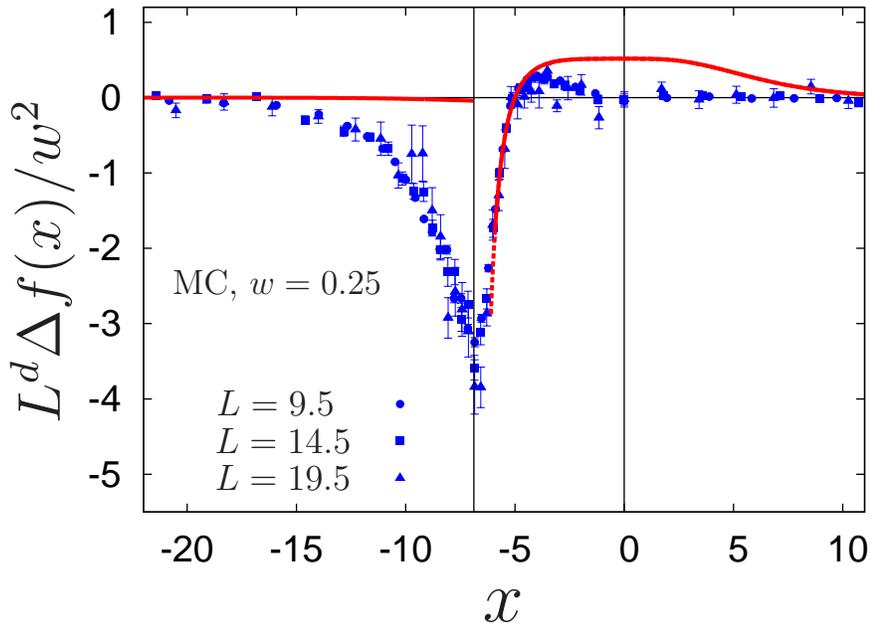}
\caption{  
The scaling function $\Delta f L^{3}/w^{2}$ of the contribution to the critical Casimir force
due to random surface fields calculated within Gaussian approximation  in $d=3$ and given by 
Eqs.~(\ref{33}) and (\ref{34}). These results are compared with  Monte Carlo (MC) simulations 
data or the $3d$ Ising film with 
random surface fields taken from Fig.~3(b) in Ref.~\cite{MVDD}. 
In order to obtain the best fit, the analytic  results 
have been rescaled as follows: 
 $x \to 1.55 x$, $y \to 0.3 y$. For the $3d$ Ising model 
 the   random surface field  scaling variable   is $\hat w=h/L^{0.26}$ 
(see   Eq.~(\ref{eq:w})). 
The vertical lines indicate $T_{c, bulk}$ and $T_{c, film}(L)$ within  MFT, 
 corresponding to $x=0$ and $x=-6.88$, respectively.
Concerning the reason for the discrepancy between the MC data and the 
analytic result below $x \lesssim -6.88$ see the main text after Eq.~(\ref{34}).
}\label{fig:2}
\end{figure}
%


\vspace{5mm}

  Analogous calculations (see Appendix B) for  the contribution of random surface fields to the critical Casimir force
in the disordered film phase for $-\pi^{2}/L^{2} = \; \tau_{c} < \tau < 0$ and for $\tau > 0$ yield
\begin{equation}
 \label{34}
\Delta f  = 
-\frac{\mathcal{A}(d)w^{2}}{L^{d}}\left\{
                          \begin{array}{ll}
\int_{0}^{1}dy \frac{x_{-}^{d+1} y^{d-2} (1-y^{2})}{
2^{(d+1)/2}\cos^{2}\Bigl(\frac{x_{-}}{2\sqrt{2}}\sqrt{1-y^{2}}\Bigr)} \\
-\int_{1}^{\infty}dy \frac{x_{-}^{d+1} y^{d-2} (y^{2}-1)}{
2^{(d+1)/2}\cosh^{2}\Bigl(\frac{x_{-}}{2\sqrt{2}}\sqrt{y^{2}-1}\Bigr)}, \;  -\sqrt{2}\pi < -x_{-} \leq 0
\\
\\
-\int_{0}^{\infty}dy \frac{x_{+}^{d+1} y^{d-2} (y^{2}+1)}{
\cosh^{2}\Bigl(\frac{x_{+}}{2}\sqrt{y^{2}+1}\Bigr)}, \; \; \;   x_{+}  \geq 0
                          \end{array}
\right.
\end{equation}
where  $x_- =-L/\xi_{-}$ and  $ x_{+} = L/\xi_{+}$.
  Note that  because in the disordered  phase the mean field OP profile $\psi_{-}(s,x_{-})$ 
is identically equal to zero, the derivation of the above result turns out to be much more simple
than  the one for  the ordered  phase  in Eq.~(\ref{33}).
Whereas Eq.~( \ref{33}) is only approximately valid for $x_{-} \gg 1$, i.e., $x =-x_{-} \to -\infty$,
Eq.~(\ref{34}) holds for  $0 > -x_{-}  \gtrsim -\sqrt{2} \pi$,  i.e., not too close to $\tau_c$, and for  
$x_{+} = L/\xi_{+} \geq 0$.
The scaling function $\Delta f L^{d}/w^{2}$ of the random field contribution to the critical Casimir force
as given by Eqs. (\ref{33}) and (\ref{34})  is shown in Fig.~\ref{fig:1}.


\section{Discussion and perspectives}
\label{sec:dis}

It is interesting  and instructive to compare the qualitative behavior of the contribution to the critical Casimir force
due to random  surface fields
with the corresponding force for the pure system with Dirichlet-Dirichlet boundary conditions.
In the absence of  random   surface fields (i.e., $h=0$) the free energy
is given by the first two terms  on the r.h.s. of Eq.~(\ref{23}). There, the first term is the
standard mean field contribution  (Eq.~(\ref{11})), while the second term
stems from the Gaussian fluctuations 
described by the correlation function matrix given in Eq.~(\ref{15}).
Accordingly one finds for the CCF $f_0$ (per $k_BT$ and per area $S_{d-1}$ and in excess of the 
$L$-independent contribution from the bulk free energy)
 $f_0 = -(\partial E_0/\partial L)+f_0^{(G)} $, where $f_0^{(G)}$ is the contribution from the Gaussian fluctuations. 
(The surface  free energy of the  film  does not depend on the film thickness 
and thus it   does not render a contribution to $f_0$.)
 An  analytical expression for the mean field contribution $-\partial E_0/\partial L$ 
is available only for $d=4$; it is given by Eq.~(56) and Fig.~9 in Ref.~\cite{mgd}.   This result  
vanishes $\sim |x_{-}|^2\exp(-\sqrt{2}|x_{-}|)$ for $-x_{-}\to -\infty$, is parabolic for $-\sqrt{2}\pi \le -x_{-}\le 0$, 
and is zero for $x_{+}>0$. 
For $T<T_{c,b}$ the Gaussian contribution $f_0^{(G)}$ must be determined numerically (second term in Eq.~(\ref{23})).
For $T>T_{c,b}$ one has $f_0^{(G)}(x_{+}>0,d=4)=3\Theta_{+(0,0)}(x_{+})-x_{+}\Theta'_{+(0,0)}(x_{+})$ 
with $\Theta_{+(0,0)}(x_{+})= -  \bigl(x_{+}^4/(6\pi^2) \bigr)\int_1^{\infty}dy(y^2-1)^{3/2}/(e^{2x_{+}y}-1)$
(see Eq.~(6.12) for $\epsilon =0$ in the first entry of Ref.~\cite{dietrich_krech}); 
accordingly $f_0^{(G)}(x_{+}\to \infty)=- \bigl(1/(16\pi^{3/2})\bigr)x_{+}^{3/2}e^{-2x_{+}}$.
For $d=3$, the numerically evaluated mean field contribution is shown in Fig.~13 of Ref.~\cite{VGMD}.
For $d=3$ and  $T<T_{c,b}$, as for $d=4$,  the Gaussian contribution $f_0^{(G)}$ must be determined numerically.
For $d=3$ and  $T>T_{c,b}$ one has $f_0^{(G)}(x_{+}>0,d=3)=3\Theta_{+(0,0)}(x_{+})-x_{+}\Theta'_{+(0,0)}(x_{+})$ 
with $\Theta_{+(0,0)}(x_{+})= - x_{+}^{3/(4\pi)}\int_1^{\infty}dy(y^2-1)/(e^{2x_{+}y}-1)$
(see Eq.~(6.6)  in the first entry of Ref.~\cite{dietrich_krech}); 
accordingly $f_0^{(G)}(x_{+}\to \infty)=- \bigl(1/(6\pi)\bigr)x_{+} e^{-2x_{+}}$.

Our results obtained within the Gaussian approximation  for weak disorder in $d=3$
(Eqs.~(\ref{33})) confirm the the interpretation 
of the MC simulation data in Ref.~\cite{MVDD}, formulated therein as a hypothesis. 
This hypothesis  states  that for small values of $ w$ the contribution $\Delta f$ to the critical Casimir force
due to random  surface fields is, to  leading order,  proportional to $w^2$, i.e., 
for the scaling function $\vartheta$ of the critical Casimir force one has
\begin{equation}
\label{eq:scal_force_w}
 f_0(T,L,h) L^{3} = \vartheta(  x,w)\approx \vartheta( x , w=0) +  w^2 \delta\vartheta(x) \, ,
\end{equation}
where $\vartheta( x,w=0)$ is the scaling function of the critical Casimir force for $(o,o)$ BC without RSF and
 the universal scaling
function  $\delta\vartheta$, which is defined via Eq.~(\ref{eq:scal_force_w}),  depends on $x$ only. 
  The scaling variable $x$ equals  $-L/\xi_{-}$ for $T \leq T_{c,b}$ and $+L/\xi_{+}$ for $T \geq T_{c,b}$.
In Fig.~\ref{fig:2}, we compare  
$\Delta f L^{d}/w^{2} = L^d(f_0(T,L,h) - f_0(T,L,h=0))/w^2 \; \simeq \delta\vartheta(x) $ as given by Eqs.~(\ref{33}) and (\ref{34})
for $d=3$  with the MC simulation data obtained  in Ref.~\cite{MVDD} 
for  $3d$ Ising films with weak surface disorder corresponding to the scaling variable $\hat w=h/L^{0.26}=0.25$.
(In the Ising model considered in Ref.~\cite{MVDD}, the coupling constant within  the surface layers 
and between the surface layers and their  neighboring layers has been taken to be the same as in the bulk.
The  corresponding surface enhancement is, within  mean-field theory and in units
of the lattice spacing,
$c=1$ \cite{binder}. Beyond mean field theory, the relation between  $c$ and 
the coupling constants is not known. In Ref.~\cite{MVDD}
the value of $c$ has been  set such that  $c^{0.87}=1/\kappa$ and   
 the scaling variable $\hat w=h/L^{0.26}$ has been used.) 
 The best fit of the MC data by the analytical result is achieved by stretching and compressing 
  the scaling variable $x$ and the amplitude
of the analytic result for $\Delta f L^{d}/w^{2}$ 
by a factor  of $1.55$ and of $0.3$, respectively.
As can be inferred from Fig.~\ref{fig:2},
the Gaussian approximation qualitatively captures 
the influence  of the random surface fields on the CCF in the case of weak disorder.
Quantitative agreement is not expected
 and, indeed, we find that for $-15\lesssim x \lesssim 10$ the analytic result for $d=3$ deviates from the MC data.
The observed discrepancy is enhanced by the fact that the analytic calculations have been performed  by assuming the limit
 $c \to \infty$,  whereas the MC simulation data have been obtained for $c\simeq 1$.
Moreover, for $x< x_c = -\sqrt{2}\pi$, the scaling function for the OP profile has been  approximated by the
scaling function  for the associated semi-infinite system  close to its fixed-point form corresponding to $c=\infty$
(compare Eqs.~(\ref{16a}) and (\ref{17})).
As already discussed earlier (see Section II, Eqs.~(\ref{16a}) and (\ref{17})), 
 this approximation is valid for $x_- \gg 1$. 
As can be seen in Fig.~\ref{fig:3}, for $c=1$, which corresponds to the model system studied within the MC simulation,
even for $x_-$ as large as 20 the deviation of the  OP scaling function for a film  from the one for the  corresponding semi-infinite
system is considerable. The  smaller the film thickness, the stronger is the deviation.

Concerning future studies, it would be desirable to consider  spatially correlated random surface fields with nonzero mean 
which better mimic the actual physical systems. 
In addition, it would be interesting to study to which extent random surface fields eliminate the critical point
$T_{c, film}$ of the film and, if not, how $T_{c, film}$ is shifted by the Gaussian fluctuations with and without
random surface fields.

Finally it would be rewarding to make analytic progress beyond the Gaussian approximation. To this end one can extend 
the renormalization group analysis for the energy density at a single surface (i.e., for a semi-infinite 
system \cite{DD}) to that in the presence of a second surface at a distance $L$ (i.e., for the film geometry).
This will lead to a scaling form of the surface energy density which is complicated due to the combination
of multiplicative and additive renormalization. Even the comparison of this scaling property with the present} 
explicit Gaussian result is expected to be impeded by logarithmic corrections
appearing in $d=4$. Moreover, in order to be consistent the relation in Eq.~(\ref{25}) has to be 
augmented in order to capture non-Gaussian contributions.

\begin{figure}[h]
  \includegraphics[width=0.5\textwidth]{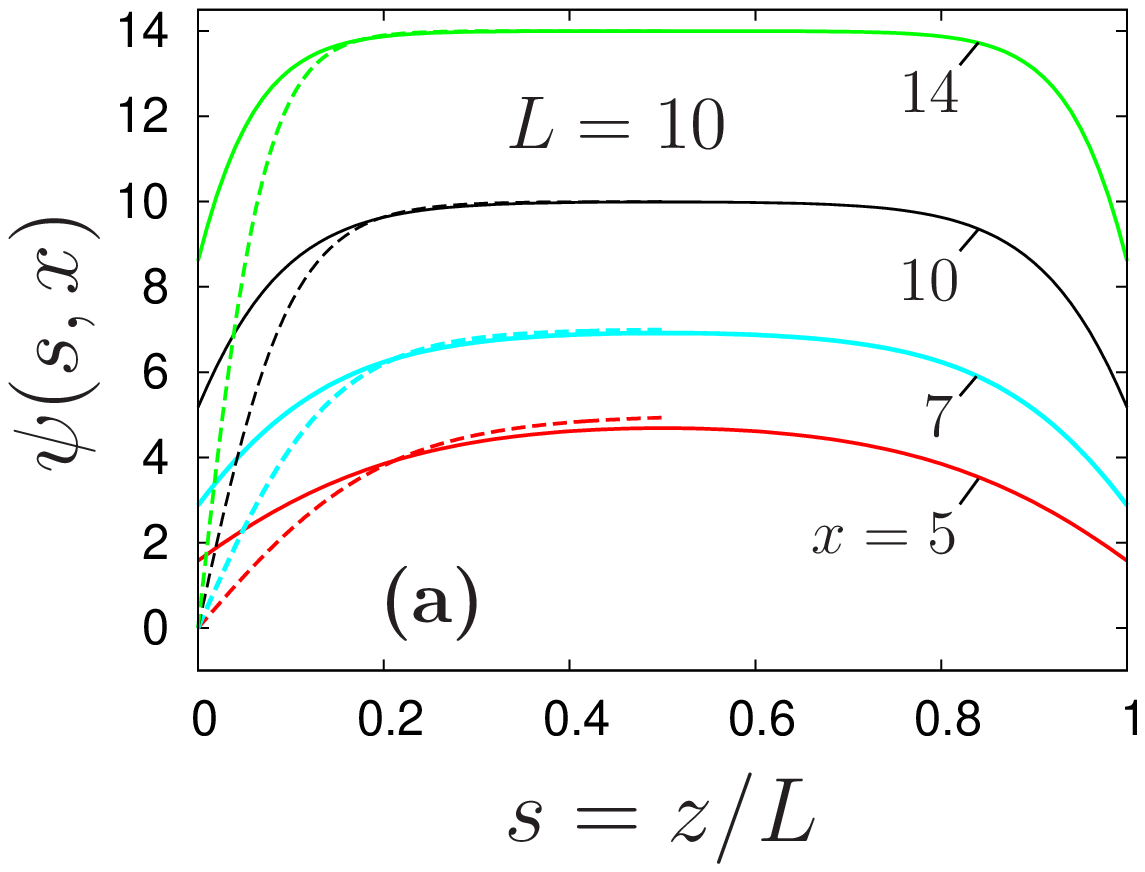}
   \includegraphics[width=0.5\textwidth]{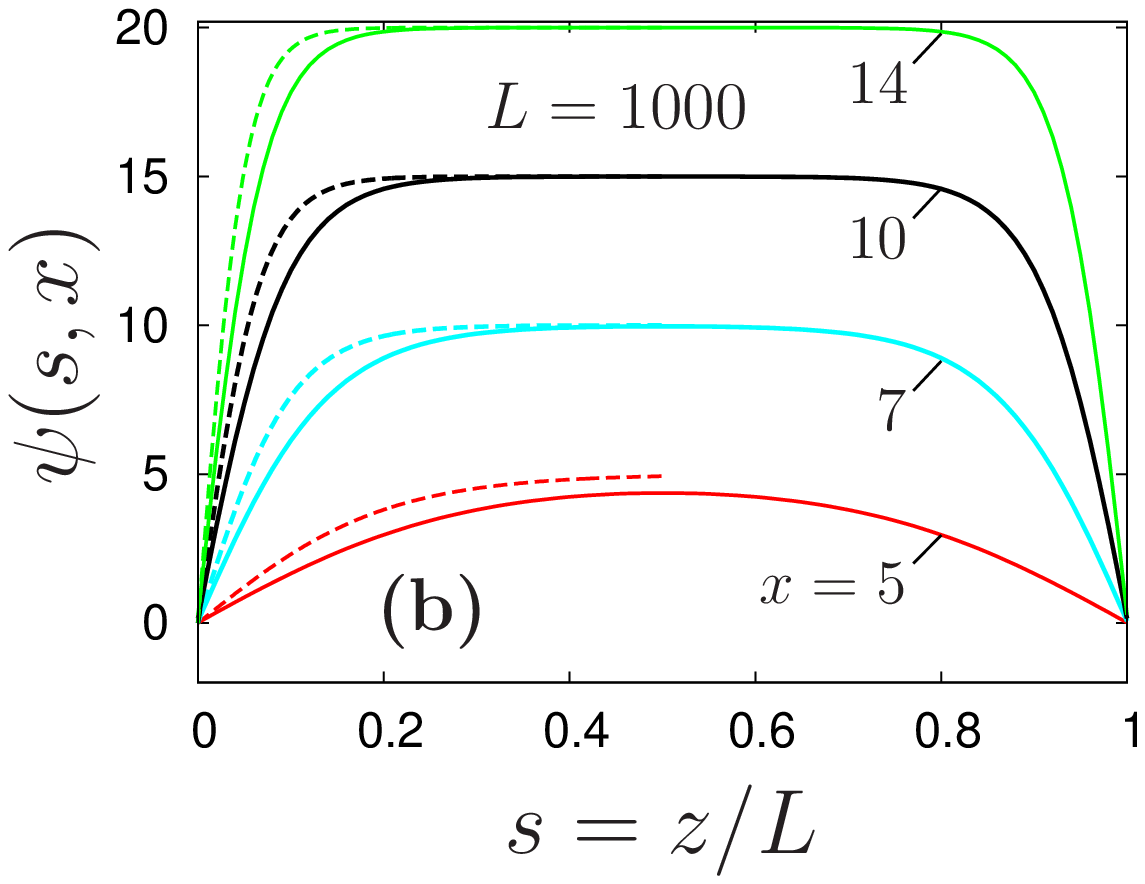}
\caption{Scaling function $\psi(s,x)$ of the OP profile in the film (see Sec. III) of thickness $L=10$ 
(a) and $L=1000$ (b)  with  $(o,o)$ BCs 
for various values of the scaling variable $x_-=L/\xi_-$ (solid lines)
as obtained within  mean field theory for  $c=1$ (see Ref.~\cite{Gambassi_Dietrich})
 compared with the ones for the corresponding semi-infinite system at the fixed point $c=\infty$ (dashed lines).
 For the  narrow film with $L=10$, deviations from the semi-infinite
OP scaling function are more pronounced and occur  at smaller values of $x_-$ as for the thick films.
Such narrow films have been  studied by the MC simulations reported in Ref.~\cite{MVDD}.
}
\label{fig:3}
\end{figure}

\textbf{Acknowledgments:} The work by AM has been supported by the Polish National Science Center 
(Harmonia Grant No. 2015/18/M/ST3/00403). 
\newpage

\begin{center}

\appendix{\Large \bf Appendix A: ordered phase in the film}

\end{center}

\newcounter{A}
\setcounter{equation}{0}
\renewcommand{\theequation}{A\arabic{equation}}

\vspace{5mm}

In this appendix we consider the Gaussian fluctuations around a nonzero mean field order parameter $\phi_{*} \not= 0$.
This occurs at $\tau < \tau_{c} = -\pi^{2}/L^{2}$, i.e., below the bulk critical point \cite{Gambassi_Dietrich}.
In terms of the scaling variable $x(\tau < 0) = -x_{-}$ this appendix is concerned with $x < \sqrt{2} \pi$. 

Accordingly, due to to Eqs.~(\ref{15}) and (\ref{17a}) the matrix  elements $G_{l,l'}^{-1}$ are given as
\begin{equation}
 \label{A1}
 G_{l,l'}^{-1} \; = \; a_{l} \delta_{l,-l'} - \tilde m(l+l'; x_{-}) + 2c
\end{equation}
where
\begin{equation}
 \label{A2}
 a_{l} = L \Bigl[p^{2} +\xi_{-}^{-2} + \Bigl(\frac{2\pi l}{L}\Bigr)^{2} \Bigr]
\end{equation}
and  approximately
\begin{equation}
 \label{A3}
\tilde m(l; x_{-}) = \frac{3}{\xi_{-}} \int_{0}^{x_{-}/2} ds \; \cosh^{-2}(s) \cos\Bigl( \frac{2\pi l}{x_{-}} s\Bigr)
\end{equation}
In view of Eq.~(\ref{25})  our aim is to compute the quantity
\begin{equation}
 \label{A4}
 S \; = \; \sum_{l,l'=-\infty}^{+\infty}G_{l,l'}
\end{equation}
where the matrix $\hat{G} = \bigl(G_{l,l'}\bigr)$ is the inverse of the matrix $\hat{G}^{-1} = \bigl(G_{l,l'}^{-1}\bigr)$ given by Eq.~(\ref{A1}). 
It will turn out that the above sum $S$ can be computed without making use of an  explicit   expression for the matrix elements $G_{l,l'}$.

To start with, we consider the matrices  $G_{l,l'}$ and $G_{l,l'}^{-1}$  to have a very large
but finite rank $N\times N$; only in the final result we shall take the limit $N\to\infty$.
By definition the inverse matrix fulfills
\begin{equation}
 \label{A5}
 \sum_{l'=-N}^{N} G_{l_{1},l'}^{-1} \, G_{l',l_{2}} \; = \; \delta_{l_{1}, l_{2}} \; .
\end{equation}
Summing the above relation over $l_{1}$ and $l_{2}$ we find
\begin{equation}
 \label{A6}
 \sum_{l=-N}^{N} \tilde{C}_{l} \, C_{l} \; = \; 2N +1
\end{equation}
where
\begin{equation}
 \label{A7}
 C_{l} \; = \; \sum_{l'=-N}^{N} G_{l,l'} \; 
\end{equation}
and, according to Eq.~(\ref{A1}),
\begin{eqnarray}
 \nonumber
 \tilde{C}_{l}  =  \sum_{l'=-N}^{N} G_{l,l'}^{-1}  & = &
                   \sum_{l'=-N}^{N} \Bigl[ a_{l} \delta_{l,-l'} - \tilde m(l+l';x_{-}) + 2c\Bigr]
\\
\nonumber
\\
&=&
a_{l} - M(x_{-},l) +2c (2N+1)
 \label{A8}
\end{eqnarray}
where 
\begin{equation}
 \label{A9}
 M(x_{-},l) \; = \; \sum_{l'=-N}^{N} \tilde m(l+l'; x_{-}) \; = \; 
 \frac{3}{\xi_{-}} \sum_{l''=-N+l}^{N+l} \int_{0}^{x_{-}/2} ds \; \cosh^{-2}(s) \cos\Bigl( \frac{2\pi l''}{x_{-}} s\Bigr).
\end{equation}
$M(x_{-},l)$ can be written as 
\begin{eqnarray}
 \label{A9a}
 M(x_{-},l) \; &=& M(x_{-}) + R(l;x_{-}) \; = \; \sum_{l'=-N}^{N} \tilde m(l'; x_{-}) + R(l; N;x_{-})
 \nonumber \\
\; &=& \;\frac{3}{\xi_{-}} \sum_{l''=-N}^{N} \int_{0}^{x_{-}/2} ds \; \cosh^{-2}(s) \cos\Bigl( \frac{2\pi l''}{x_{-}} s\Bigr)
 + R(l; N;x_{-})
\end{eqnarray}
 with $M(x_{-}) = M(x_{-}, l=0)$. 
In the limit of large $N$ (which will be taken to infinity  in the final result) 
one has $R(l;N\to\infty; x_{-})=0$ for all $l$ and $x_{-}$.
Substituting Eq.~(\ref{A8}) into Eq.~(\ref{A6}) and taking into account 
that according to the definition in Eq.~(\ref{A4}), one has
$S = \sum_{l} C_{l}$ so that
\begin{equation}
 \label{A10}
 \sum_{l=-N}^{N} \Bigl( a_{l}-R(l; N;x_-)\Bigr) C_{l} + \Bigl((4N+2)c - M(x_{-})\Bigr) \, S \; = \; 2N+1.
\end{equation}
This equation is satisfied by
\begin{equation}
 \label{A11}
 C_{l} \; = \; \Bigl(a_{l}-R(l; N; x_-)\Bigr)^{-1} \Bigl[ 1 + \tilde{m}(l; x_{-}) S - 2c S\Bigr] \; .
\end{equation}
Summing Eq.~(\ref{A11}) over $l$ we obtain a simple equation for $S$:
\begin{equation}
 \label{A12}
 S \; = \; \sum_{l=-N}^{N} \Bigl(a_{l}-R(l;N;x_{-})\Bigr)^{-1} \; + \;
            S \sum_{l=-N}^{N} \frac{\tilde m(l; x_{-})}{a_{l}-R(l;N;x_{-})} \; - \; 
2cS \sum_{l=-N}^{N} \frac{1}{a_{l}-R(l;N;x_{-})} \, .
\end{equation}
In the limit $N\to \infty$ we eventually find
\begin{equation}
 \label{A13}
 \sum_{l,l'=-\infty}^{+\infty}G_{l,l'} \; \equiv \; S \; = \; \frac{g}{1 \; + \; 2c \, g \; - g_{1}}
\end{equation}
which is Eq.~(\ref{26}). The series
\begin{equation}
 \label{A14}
 g  \; = \; \sum_{l=-\infty}^{+\infty} a_{l}^{-1}
\end{equation}
and
\begin{equation}
 \label{A15}
 g_{1}  \; = \; \sum_{l=-\infty}^{+\infty} a_{l}^{-1} \; \tilde m(l; x_{-})
\end{equation}
are still to be calculated.

With Eq.~(\ref{A2}) the series in Eq.~(\ref{A14}) can be written as
\begin{equation}
 \label{A16}
 g \; = \; \frac{L}{4\pi^{2}} \Bigl[ 2 \sum_{l=1}^{\infty} \frac{1}{l^{2} + \gamma^{2}} + \frac{1}{\gamma^{2}}\Bigr] \;  , 
\end{equation}
where
\begin{equation}
 \label{A17}
 \gamma^{2} = \frac{L^{2}}{4\pi^{2}} \bigl(p^{2} + \xi_{-}^{-2}\bigr) \; = \; \frac{x_{-}^{2}}{4\pi^{2}} \; \bigl[(\xi_{-} p)^{2} + 1\bigr]  .
\end{equation}
The series  in Eq.~(\ref{A16}) is  known as
$\sum_{l=1}^{\infty}(l^{2} + \gamma^{2})^{-1} = \frac{\pi}{2\gamma} [\tanh(\pi\gamma)]^{-1} - \frac{1}{2\gamma^{2}}$
(see Eq.~(1.217.1) in Ref. \cite{Grandshteyn}). Thus we obtain
\begin{equation}
 \label{A18}
 g \; \equiv \; g(p,L,\xi_{-}) \; =
 \; \frac{1}{2\sqrt{p^{2} + \xi^{-2}_{-}} \tanh\Bigl(\frac{L}{2} \sqrt{p^{2} + \xi_{-}^{-2}} \Bigr)} \; ,
\end{equation}
which is Eq.~(\ref{27}).

The  series in Eq.~(\ref{A15})  can be written as 
\begin{equation}
 \label{A19}
 g_{1} \; = \; \frac{3L}{4\pi^{2} \xi_{-}}    \int_{0}^{x_{-}/2} ds \; \cosh^{-2}(s) 
 \sum_{l=-\infty}^{+\infty} 
 \frac{1}{l^{2} + \gamma^{2}} \cos\Bigl( \frac{2\pi l}{x_{-}} s\Bigr) \, .
 \end{equation}
For large values of $x_{-}$ the  series in Eq.~(\ref{A19}) can be approximated by the integral
\begin{eqnarray}
 \nonumber
 g_{1} &=& \frac{3}{4\pi^{2}}    \int_{0}^{x_{-}/2} ds \; \cosh^{-2}(s) 
 \frac{1}{x_{-}} \sum_{l=-\infty}^{+\infty} 
 \frac{1}{\Bigl(\frac{l}{x_{-}}\Bigr)^{2} + \frac{1}{4\pi^{2}} \; \bigl[(p\xi_{-})^{2} + 1\bigr] } \cos\Bigl( 2\pi s\frac{l}{x_{-}}\Bigr) 
\\
\nonumber
\\
&\simeq&
\frac{3}{4\pi^{2}}    \int_{0}^{x_{-}/2} ds \; \cosh^{-2}(s) \; \int_{-\infty}^{+\infty} dt
\frac{1}{t^{2} + \frac{1}{4\pi^{2}} \; \bigl[(p\xi_{-})^{2} + 1\bigr] } \cos\Bigl( 2\pi s\; t\Bigr). 
\label{A21}
 \end{eqnarray}
Simple integration over $t$ yields
\begin{equation}
 \label{A22}
 g_{1} \equiv g_{1}\bigl( (p\xi_{-})^{2} , x_{-}\bigr) \; = \; 
 \frac{3}{2\sqrt{(p\xi_{-})^{2} + 1} } \int_{0}^{x_{-}/2} ds \; \cosh^{-2}(s) \; \exp\Bigl\{- s \sqrt{(p\xi_{-})^{2} + 1} \Bigr\}.
\end{equation}

\vspace{10mm}

\begin{center}

\appendix{\Large \bf Appendix B: disordered phase in the film}

\end{center}

\newcounter{B}
\setcounter{equation}{0}
\renewcommand{\theequation}{B\arabic{equation}}

\vspace{5mm}

{\large \bf B1: $-\pi^{2}/L^{2} < \tau \leq 0$}

\vspace{5mm}

In the disordered phase in the film below $T_{c,bulk}$ (for which the mean field equilibrium profile 
is identically zero, i.e., $\phi_{*}(z) \equiv 0$ as for  $T > T_{c,bulk}$)
the Hamiltonian, which describes the fluctuating field $\varphi({\bf r},z)$
within the Gaussian approximation, is given by 
\begin{equation}
 \label{B1}
{\cal H}_{0}[\varphi] = 
+ \frac{1}{2} \int d^{d-1}r \int_{0}^{L}dz \Bigl[\bigl(\nabla \varphi\bigr)^{2}
-\frac{1}{2} \xi_{-}^{-2}\varphi^{2} 
+ c  \varphi^{2} \bigl[\delta(z) + \delta(z-L)\bigr] \Bigr] \; .
\end{equation}
 With $-\frac{1}{2}\xi_{-}^{-2} = \tau \;$ Eq.~(\ref{B1}) holds for the interval $-\pi^{2}/L^{2} < \tau \leq 0$
in which the bulk is ordered but the film is disordered.
Inserting the Fourier representation (Eq.~(\ref{12})) into Eq.~(\ref{B1}) yields
\begin{equation}
  \label{B2}
{\cal H}_{0}[\tilde{\varphi}]   = 
 \frac{1}{2} \int\frac{d^{d-1}p}{(2\pi)^{d-1}} \, \frac{1}{L^{2}} \sum_{l,l'=-\infty}^{+\infty}
G^{-1}_{l,l'}(p) \,
\tilde{\varphi}({\bf p}, l) \tilde{\varphi}(-{\bf p}, l')
\end{equation}
 where the matrix elements $G^{-1}_{l,l'}(p)$ have  a much more simple structure compared with the ones in Eq.~(\ref{15}):
\begin{equation}
 \label{B3}
G^{-1}_{l,l'}(p) \; = \; L \Bigl[p^{2} -\frac{1}{2} \xi_{-}^{-2} + \frac{4\pi^{2}}{L^{2}} l^{2} \Bigr] \delta_{l, -l'}
                      \, + \, 2 c .
\end{equation}

Following the same steps as in the calculation for the ordered phase, for the free energy contribution $\Delta F(h,L)$,
caused by the random fields, one obtains the analogue of Eq.~(\ref{25})  for which instead of Eq.~(\ref{26}) 
one now finds a much more simple expression:
\begin{equation}
 \label{B4}
 \sum_{l,l'=-\infty}^{+\infty} G_{l,l'}(p) \; = \;
\frac{g(p,L,\xi_{-})}{
1 + 2c \, g(p,L,\xi_{-}) }
\end{equation}
with
\begin{equation}
\label{B5}
g(p,L,\xi_{-}) \; = \; \frac{1}{L}
\sum_{l=-\infty}^{+\infty} \Bigl[p^{2} - \frac{1}{2} \xi_{-}^{-2} + \Bigl(\frac{2\pi l}{L}\Bigr)^{2}\Bigr]^{-1}. 
\end{equation}
Repeating the calculations carried out  in Appendix A, which lead to the result in Eq.~(\ref{A18}), 
one finds for the domain $p^{2} > \frac{1}{2} \xi_{-}^{-2}$
\begin{equation}
\label{B6}
g(p,L,\xi_{-})\Big|_{p^{2} > \frac{1}{2} \xi_{-}^{-2}} \; \equiv \; g_{+}(p,L,\xi_{-}) \; = \;
\frac{1}{2\sqrt{p^{2}-\frac{1}{2}\xi_{-}^{-2}} \, \tanh\Bigl(\frac{L}{2} \sqrt{p^{2}-\frac{1}{2} \xi_{-}^{-2}}\Bigr)}  \; .
\end{equation}
Similar calculations for the domain $p^{2} < \frac{1}{2} \xi_{-}^{-2}$ yield
\begin{equation}
\label{B7}
g(p,L,\xi_{-})\Big|_{p^{2} < \frac{1}{2} \xi_{-}^{-2}} \; \equiv \; g_{-}(p,L,\xi_{-}) \; = \;
-\frac{1}{2\sqrt{\frac{1}{2}\xi_{-}^{-2}- p^{2}} \, \tan\Bigl(\frac{L}{2} \sqrt{\frac{1}{2} \xi_{-}^{-2}-p^{2}}\Bigr)} \; .
\end{equation}
Upon inserting Eq.~(\ref{B4}) into Eq.~(\ref{25}) and subtracting $L$-independent terms,
for large $c$, i.e., to leading order in an  expansion in terms of $1/c$,
we obtain for the corresponding excess free energy (denoted  as $\Delta\tilde{F}$)
\begin{equation}
 \label{B8}
\frac{\Delta \tilde{F}}{S_{d-1}}  \; = \; 
\frac{h^{2}}{4c^{2}}
\Biggl[
\int_{|{\bf p}|<\frac{1}{\sqrt{2} \, \xi_{-}}} \frac{d^{d-1}p}{(2\pi)^{d-1}} \; 
\frac{1}{g_{-}(p,L,\xi_{-})} + 
\int_{|{\bf p}|>\frac{1}{\sqrt{2} \, \xi_{-}}} \frac{d^{d-1}p}{(2\pi)^{d-1}} \; 
\frac{1}{g_{+}(p,L,\xi_{-})} 
\Biggr] \, .
\end{equation}
Substituting here Eqs.~(\ref{B6}) and (\ref{B7}) respectively, 
changing the integration variable according to $p = y/\sqrt{2}\xi_{-}$, 
and integrating over the angular part of the momenta we obtain
\begin{eqnarray}
\nonumber
\frac{\Delta \tilde{F}}{S_{d-1}} &=& 
\frac{ \pi^{\frac{1-d}{2}}h^{2} }{2 \Gamma\bigl(\frac{d+1}{2}\bigr) c^{2}(\sqrt{2}\xi_{-})^{d}} 
\Biggl[
- \int_{0}^{1} dy 
\sqrt{1- y^{2}}  \tan\Bigl(\frac{x_{-}}{2\sqrt{2}} \sqrt{1-y^{2}}\Bigr) \; + 
\\
\nonumber
\\
&+&
\int_{1}^{\infty} dy 
\sqrt{y^{2}-1}  \tanh\Bigl(\frac{x_{-}}{2\sqrt{2}} \sqrt{y^{2}-1}\Bigr) 
\Biggr] \, .
\label{B9}
\end{eqnarray}
Taking the negative derivative of this expression with respect to $L$,  
$-\frac{\partial}{\partial L} =  -\xi_{-}^{-1} \frac{\partial}{\partial x_{-}}$
renders the critical Casimir force $\Delta f$, per $k_{B}T$ and per area $S_{d-1}$,
in excess to its value without  random fields:
\begin{equation}
 \label{B10}
\Delta f  \,  = \,
-\frac{ \pi^{\frac{1-d}{2}} }{4\Gamma\bigl(\frac{d+1}{2}\bigr) } \;
\frac{w^{2}x_{-}^{d+1} }{L^{d}2^{(d+1)/2}}  \; 
\Biggl[                     
\int_{0}^{1}dy \frac{y^{d-2} (1-y^{2})}{
\cos^{2}\Bigl(\frac{x_{-}}{2\sqrt{2}}\sqrt{1-y^{2}}\Bigr)} \; - \; 
\int_{1}^{\infty}dy \frac{ y^{d-2} (y^{2}-1)}{
\cosh^{2}\Bigl(\frac{x_{-}}{2\sqrt{2}}\sqrt{y^{2}-1}\Bigr)} 
\Biggr] \; , 
\end{equation}
which is valid for $-\pi^{2}/L^{2} < \tau \leq 0$ or equivalently for  $-\sqrt{2} \, \pi < -x_{-} \leq 0$.

\vspace{5mm}

{\large \bf B2: $\tau > 0$}

\vspace{5mm}

For $\tau > 0$ the Gaussian Hamiltonian for the fluctuating fields is (with $\xi_{+}^{-2} = \tau$)
\begin{equation}
 \label{B11}
{\cal H}_{0}[\varphi] = 
+ \frac{1}{2} \int d^{d-1}r \int_{0}^{L}dz \Bigl[\bigl(\nabla \varphi\bigr)^{2} + \xi_{+}^{-2}\varphi^{2} 
+ c  \varphi^{2} \bigl[\delta(z) + \delta(z-L)\bigr] \Bigr] \; .
\end{equation}
Correspondingly, in the  Fourier representation (Eq.~(\ref{12})) one obtains
\begin{equation}
  \label{B12}
{\cal H}_{0}[\tilde{\varphi}]   = 
 \frac{1}{2} \int\frac{d^{d-1}p}{(2\pi)^{d-1}} \, \frac{1}{L^{2}} \sum_{l,l'=-\infty}^{+\infty}
G^{-1}_{l,l'}(p) \,
\tilde{\varphi}({\bf p}, l) \tilde{\varphi}(-{\bf p}, l')
\end{equation}
 where 
\begin{equation}
 \label{B13}
G^{-1}_{l,l'}(p) \; = \; L \Bigl[p^{2} + \xi_{+}^{-2} + \frac{4\pi^{2}}{L^{2}} l^{2} \Bigr] \delta_{l, -l'}
                      \, + \, 2 c \; .
\end{equation}
Following the same steps as above, for the free energy contribution $\Delta F(h,L)$
due to the random fields one finds Eq.~(\ref{25}) where 
\begin{equation}
 \label{B14}
 \sum_{l,l'=-\infty}^{+\infty} G_{l,l'}(p) \; = \;
\frac{g(p,L,\xi_{+})}{
1 + 2c \, g(p,L,\xi_{+}) }
\end{equation}
with
\begin{equation}
\label{B15}
g(p,L,\xi_{+}) \; = \; \frac{1}{L}
\sum_{l=-\infty}^{+\infty} \Bigl[p^{2} + \xi_{+}^{-2} + \Bigl(\frac{2\pi l}{L}\Bigr)^{2}\Bigr]^{-1}  \; = \; 
\frac{1}{2\sqrt{p^{2}+\xi_{+}^{-2}} \, \tanh\Bigl(\frac{L}{2} \sqrt{p^{2} + \xi_{+}^{-2}}\Bigr)} \; .
\end{equation}
 For $\tau > 0$, this yields the expression analogous to
Eq.~(\ref{B10}) for the  critical Casimir force in excess to its value without random fields: 
\begin{equation}
 \label{B16}
\Delta f  \,  = \,
\frac{ \pi^{\frac{1-d}{2}} }{4\Gamma\bigl(\frac{d+1}{2}\bigr) } \;
\frac{w^{2}x_{+}^{d+1} }{L^{d}}  \; 
\int_{0}^{\infty}dy \frac{ y^{d-2} (y^{2}+1)}{
\cosh^{2}\Bigl(\frac{x_{+}}{2}\sqrt{y^{2}+1}\Bigr)} 
 \; , 
\end{equation}
which is valid for $x_{+} > 0 $.

\end{document}